\documentclass{aa}
\usepackage{graphicx}
\usepackage{rotating}
\usepackage{txfonts}
\begin{document}

\title{Decoding of the light changes in eclipsing Wolf-Rayet binaries\\
I.\ A non-classical approach to the solution of light curves}
\author{C.\ Perrier\inst{1} \and J.\ Breysacher\inst{2} \and G.\ Rauw\inst{3}\thanks{Also Research Associate FNRS (Belgium)}}
\institute{Observatoire de Grenoble, Universit\'e Joseph Fourier, F-38041 Grenoble Cedex, France \and Lieu-dit Petit Riston,F-40230 Saubion, France \and Institut d'Astrophysique et de G\'eophysique, Universit\'e de Li\`ege, All\'ee du 6 Ao\^ut, B\^at B5c, B-4000 Li\`ege (Sart Tilman), Belgium}
\mail{rauw@astro.ulg.ac.be}
\authorrunning{C.\ Perrier et al.}
\titlerunning{Decoding the light changes of eclipsing WR binaries}

\abstract{}{We present a technique to determine the orbital and physical parameters of eclipsing eccentric Wolf-Rayet + O-star binaries, where one eclipse is produced by the absorption of the O-star light by the stellar wind of the W-R star.}{Our method is based on the use of the empirical moments of the light curve that are integral transforms evaluated from the observed light curves. The optical depth along the line of sight and the limb darkening of the W-R star are modelled by simple mathematical functions, and we derive analytical expressions for the moments of the light curve as a function of the orbital parameters and the key parameters of the transparency and limb-darkening functions. These analytical expressions are then inverted in order to derive the values of the orbital inclination, the stellar radii, the fractional luminosities, and the parameters of the wind transparency and limb-darkening laws.}{The method is applied to the SMC W-R eclipsing binary HD\,5980, a remarkable object that underwent an LBV-like event in August 1994. The analysis refers to the pre-outburst observational data. A synthetic light curve based on the elements derived for the system allows a quality assessment of the results obtained.}{} 
\maketitle

\section{Introduction}
Photometric monitoring of Wolf-Rayet (W-R) binaries revealed that many of them display a shallow eclipse when the W-R star passes in front of its O-type companion (e.g.\ Lamontagne et al.\ \cite{lamontagne}). These so-called atmospheric eclipses arise when part of the light of the O-type companion is absorbed by the wind of the W-R star. In a few cases, the light curve displays an eclipse at both conjunctions and the analysis of this phenomenon can provide important information about the physical parameters of W-R stars and their winds. In this context, the most famous example is V444\,Cyg (WN5 + O6\,V), which has been extensively investigated by the Moscow group (e.g.\ Antokhin \& Cherepashchuk \cite{Igor} and references therein). Cherepashchuk and coworkers (e.g.\ Cherepashchuk \cite{Cher75}, Antokhin \& Cherepashchuk \cite{Igor} and references therein) developed a sophisticated method to handle the ill-posed problem of light curve inversion for V444\,Cyg. Based on the minimum {\it a priori} assumptions about the transparency function, this method not only yields the radii of both components, but also provides information about the structure of the WN5's stellar wind. However, because of a number of fundamental hypotheses that are not necessarily valid for all eclipsing W-R binaries, this method cannot be readily applied to all eclipsing W-R + O systems. For eccentric systems in particular (e.g.\ WR22, Gosset et al.\ \cite{wr22}) some assumptions (such as spherical symmetry of the problem) break down and a different technique must be used. 

We initiated our study of the observed light changes of eclipsing Wolf-Rayet binary systems when two of us (J.B. and C.P.) tried to confirm the 25.56 day period found by Hoffmann et al.\ (\cite{Hoffmann}) for the SMC star HD\,5980, the first extragalactic Wolf-Rayet binary then known to display eclipses. This exercise led to the discovery of the correct orbital period of HD\,5980, $P = 19.266 \pm 0.003$ days (Breysacher \& Perrier \cite{BP80}), the light curve revealing, in addition, a rather eccentric orbit of $e = 0.47$ assuming $i = 80^{\circ}$. However, because of the uncertainties in the depth of both minima, caused by an insufficient number of observations, no detailed quantitative analysis of this preliminary light curve could be attempted.

Its relatively long period and large eccentricity ensure that HD\,5980 is an interesting object in which to study the structure of a W-R envelope, and a photometric monitoring of this system was initiated to define the shape of its light curve in a more accurate way. More than 700 observations were collected. After realizing that none of the existing `classical tools' was suited to our purpose -- the decoding of the light changes of a {\it partially-eclipsing system} characterized by an {\it eccentric orbit} and containing one component with an {\it extended atmosphere} -- we started to develop another approach to the solution of light curves. 

The technique of light curve analysis applied to V444\,Cyg by Smith \& Theokas (\cite{ST}), which is based on Kopal's fundamental work (cf.\ Kopal \cite{Kop1}, \cite{Kop2}), appeared as an attractive approach to the solution of our problem. This method is based on the interpretation of the observed light changes in the {\it frequency-domain}, i.e., not the light curve as a function of time, but its {\it Fourier-like integral transform}.

We now describe in detail the method we developed for the study of Wolf-Rayet eclipsing binaries. A preliminary application of this technique to the light curve of HD\,5980 prior to its 1994 LBV-like eruption (see e.g., Bateson \& Jones \cite{BJ}, Barb\'a et al.\ \cite{Barba}, Heydari-Malayeri et al.\ \cite{HM}) was presented by Breysacher \& Perrier (\cite{BP91}, hereafter BP91). We reanalyse the pre-outburst observational data using an improved version of the software tool. Revised values for the physical
parameters of HD\,5980 are derived. A synthetic light curve based on the elements thus obtained allows a quality assessment of the new results.

A more detailed study of HD\,5980, including the analysis of the light curve obtained after the eruption (Sterken \& Breysacher \cite{Ster97}), will be presented in a forthcoming paper.

\section{Analysis of the light changes in the frequency-domain}
In this section, we present the fundamental equations of our method, we then introduce the mathematical functions to model the transparency and limb-darkening functions and consider the specific problem of eccentric orbits.
\subsection{The basic equations}
\begin{figure}[htb]
\begin{center}
\resizebox{8cm}{6.4cm}{\includegraphics{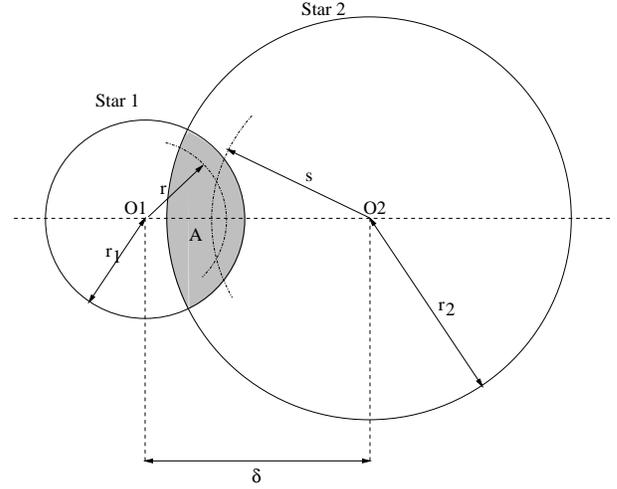}}
\end{center}
\caption{Geometry of the eclipse of star 1 by star 2. The integral of Eq.\,(\ref{eqn1}) is evaluated over the hatched area $A(r_1,r_2,\delta)$. The points inside this area can be specified by the coordinates $(r, s)$ corresponding to the two intersections of the circle with radius $r$ and centre O$_1$ and the circle with radius $s$ centred on O$_2$.\label{fig-1}}
\end{figure}

We refer to the fundamental work of the Manchester group (cf.\ Kopal \cite{Kop1}, \cite{Kop2}; Smith \cite{Smith}), and first consider an eclipsing system consisting of two spherical stars revolving around the common centre of gravity in circular orbits, and appearing in projection on the sky as uniformly bright discs. The system is seen at an inclination angle $i$. When star 1 of fractional luminosity $L_{1}$ and radius $r_{1}$ is partly eclipsed by star 2 of fractional luminosity $L_{2}$ and radius $r_{2}$ (Fig.\,\ref{fig-1}), the apparent brightness $l$ of the system (maximum light between minima taken as unity) is given by
\begin{equation}
l (r_{1},r_{2},\delta,J) = 1 - \int\int_{A} J(r)\,d\sigma \label{eqn1},
\end{equation}
where $\delta$ is the apparent separation of the centres of the two discs, $J$ represents the distribution of brightness over the apparent disc of the star undergoing eclipse, and $d\sigma$ stands for the surface element. The distances $r_1$, $r_2$, and $\delta$ are expressed in units of the orbital separation. The integral in Eq.\,(\ref{eqn1}) provides the apparent `loss of light' displayed by the system when an area $A(r_{1},r_{2},\delta)$ of star 1 is eclipsed (see Fig.\,\ref{fig-1}). The assumption that star 1 is uniformly bright yields
\begin{equation}
J(r) = \frac{L_{1}}{\pi r_{1}^{2}} \label{eqn2}.
\end{equation}
Combining Eqs.\,(\ref{eqn1}) and (\ref{eqn2}), we obtain
\begin{equation}
1 - l (r_{1},r_{2},\delta,J) = \frac{L_{1}}{\pi r_{1}^{2}} \int\int_{A} d\sigma = \alpha L_{1},
\end{equation}
where $\alpha$ is the ratio of the mutual area of eclipse to the area of the disc of the eclipsed star, and is a function of $r_{1}$, $r_{2}$, and $\delta$ (Kopal \cite{Kop1}). A generalisation of these concepts to the case of spherical stars with arbitrary limb-darkening laws $J(r)$ was presented by Smith (\cite{Smith}).

\begin{figure}[htb]
\begin{center}
\resizebox{8cm}{6.5cm}{\includegraphics{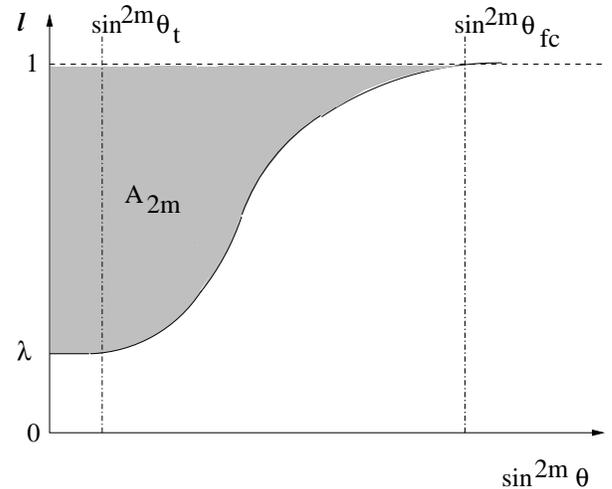}}
\end{center}
\caption{Light curve of an eclipse in the $(l, \sin^{2m}\theta)$ plane. $\theta_{\rm fc}$ corresponds to the phase angle of first contact, whilst $\theta_{\rm t}$ represents the phase angle corresponding to the beginning of the total eclipse. The shaded area illustrates the moment $A_{2m}$ and $\lambda = 1 - l(\theta = 0)$.\label{fig-2}}
\end{figure}

For an orbital period $P$ and an epoch of conjunction $t_0$, we define the phase angle $\theta$ at a time $t$ to be
\begin{equation}
\theta = \frac{2\,\pi}{P}\,(t-t_0) \label{eqn4}.
\end{equation}
As proposed by Kopal (\cite{Kop1}, \cite{Kop2}), we focus our attention on the area subtended by the light curve in the $(l, \sin^{2m}\theta)$ plane, where $m$ is a positive integer number $(m = 1, 2, 3,...)$, as shown in Fig.\,\ref{fig-2}. The areas $A_{2m}$ between the lines $l = 1$, $\sin^{2m}\theta = 0$, and the true light curve are then given by the integrals
\begin{equation}
A_{2m} = \int_{0}^{\theta_{\rm fc}} (1 - l)\,d(\sin^{2m}\theta) \label{eqn5},
\end{equation}
which are hereafter referred to as the {\it moments of the eclipse}, of index $m$, where $\theta_{\rm fc}$ denotes the phase angle of the first contact ($\delta (\theta_{\rm fc}) = r_1 + r_2$) of the eclipse.

Combining Eqs.\ (\ref{eqn1}) and (\ref{eqn5}), Kopal (\cite{Kop1},\cite{Kop2}) and Smith (\cite{Smith}) demonstrated that, based on certain assumptions, it is possible to
\begin{itemize}
\item[$\bullet$] derive analytical expressions of the moments of the eclipse in terms of the physical parameters ($i, r_1, r_2, L_1, L_2$) of a binary system with a circular orbit consisting of uniformly bright spherical stars, \item[$\bullet$] invert these relationships to determine the parameters of the system in terms of the moments $A_{2m}$ that can be empirically obtained from the data,
\item[$\bullet$] extend this treatment to the case of partial eclipses of stars with an arbitrary (yet analytical) limb-darkening law.
\end{itemize}

Going one step further, Smith \& Theokas (\cite{ST}) generalised the above concepts to derive a convenient mathematical solution to the problem of an atmospheric eclipse, i.e., an eclipse of a limb-darkened star by a star surrounded by an extended atmosphere. In such an eclipse, at each position inside the area $A$ specified by the coordinates $r$ and $s$ (see Fig.\,\ref{fig-1}), a fraction of the light emitted by star 1 is absorbed by the atmosphere of star 2. To account for these transparency effects, a transparency function $F(s)$ is introduced into Eq.\,(\ref{eqn1}), so that the total amount of light seen by the observer becomes
\begin{equation}
l (r_{1},r_{2},\delta,J,F) = 1 - \int\int_{A} J(r) F(s)\,d\sigma \label{eqn6}.
\end{equation}
Considering that in the case of an atmospheric eclipse, it might be interesting to give more weight to the data close to mid-minimum, Smith \& Theokas (\cite{ST}) also introduced an alternative set of moments $B_{2m}$ defined by
\begin{equation}
B_{2m} = - \int_{0}^{\theta_{\rm fc}} (1 - l)\,d(\cos^{2m}\theta) \label{eqn7}.
\end{equation}
The use of kernel $d(\cos^{2m}\theta)$ in Eq.\,(\ref{eqn7}) places more emphasis on the data points close to mid-eclipse, which leads to smaller errors than in the case of the $A_{2m}$ moments defined by means of the $d(\sin^{2m}\theta)$ kernel (Theokas \& Smith 1983, private communication). We define $\varepsilon$ to be the mean error in an individual data point; the relative error in the light curve $\varepsilon/(1 - l(\theta))$ increases for data points near $\theta_{\rm fc}$. While these points are given more weight by the kernel of the $A_{2m}$ moments, the converse situation holds for the $B_{2m}$ moments, where the kernel reaches its peak for a given $m$ in a zone where $1 - l(\theta)$ is closer to its maximum value.

By definition, the $A_{2m}$ and $B_{2m}$ moments are related to each other by means of 
\begin{equation}
B_{2m} = \sum^{m}_{p=1} \frac{m!}{(m-p)!\,p!}\,(-1)^{p+1}\,A_{2p} \label{eqn8}.
\end{equation}
We now concentrate on the $B_{2m}$ moments because they are equally well suited to the analysis of the primary and secondary minima with the transparency and limb-darkening functions adopted in the present study (cf.\ Section 2.2). Since there are a number of typos in the paper of Smith \& Theokas (\cite{ST}), we provide below the mathematical details of the method.  

The infinitesimal element of area $d\sigma$ of Eq.\,(\ref{eqn6}) is expressed as
\begin{equation}
d\sigma = \frac{1}{2} \frac{\partial^{2}}{\partial r \partial s} \left( \pi r^{2} \alpha(r,s,\delta)
\right) dr ds \label{eqn9}.
\end{equation}
The following general expression for the $B_{2m}$ moments was derived by Smith \& Theokas (\cite{ST})
\begin{eqnarray}
B_{2m} & = &\lambda + \int_{0}^{r_{1}} \int_{0}^{r_{2}} J(r)\,F(s)\,ds\,dr \nonumber\\
& & \times \frac{\partial^{2}}{\partial r \partial s} \left( \int_{0}^{\theta_{\rm fc}} \cos^{2m}\theta\,\frac{\partial (\pi\,r^{2}\,\alpha
(r,s,\delta))}{\partial \theta} d\theta \right),
\end{eqnarray}
where $1 - l(\theta = 0)$ is defined as $\lambda$ (see Fig.\,\ref{fig-2}).

The analytical expressions obtained for the $B_{2m}$'s for $m = 1, 2, 3, 4,$ and $5$ are thus
\begin{equation}
B_{2} = \lambda - \csc^{2}{i}\,(P - I_{1}R_{1}r_{2}^{2} - \psi_{1}),
\end{equation}
\begin{equation}
B_{4} = \lambda - \csc^{4}{i}\,(P - 2\,I_{1}R_{1}r_{2}^{2} + I_{2}R_{1}r_{1}^{2}r_{2}^{2} + I_{1}R_{2}r_{2}^{4} - \psi_{2}),
\end{equation}
\begin{eqnarray}
B_{6} & = & \lambda - \csc^{6}{i}\,(P - 3\,I_{1}R_{1}r_{2}^{2} + 3\,I_{2}R_{1}r_{1}^{2}r_{2}^{2} + 3\,I_{1}R_{2}r_{2}^{4} \nonumber \\ 
& & - I_{3}R_{1}r_{1}^{4}r_{2}^{2} - I_{1}R_{3}r_{2}^{6} - 3\,I_{2}R_{2}r_{1}^{2}r_{2}^{4} - \psi_{3}),
\end{eqnarray}
\begin{eqnarray}
B_{8} & = & \lambda - \csc^{8}{i}\,(P - 4\,I_{1}R_{1}r_{2}^{2} + 6\,I_{2}R_{1}r_{1}^{2}r_{2}^{2} + 6\,I_{1}R_{2}r_{2}^{4} \nonumber \\
& & - 4\,I_{3}R_{1}r_{1}^{4}r_{2}^{2} - 4\,I_{1}R_{3}r_{2}^{6} - 12\,I_{2}R_{2}r_{1}^{2}r_{2}^{4} + I_{4}R_{1}r_{1}^{6}r_{2}^{2} \nonumber \\
& & + I_{1}R_{4}r_{2}^{8} + 6\,I_{3}R_{2}r_{1}^{4}r_{2}^{4} +
6\,I_{2}R_{3}r_{1}^{2}r_{2}^{6} - \psi_{4}),
\end{eqnarray}
and
\begin{eqnarray}
B_{10} & = & \lambda - \csc^{10}{i}\,(P - 5\,I_{1}R_{1}r_{2}^{2} + 10\,I_{2}R_{1}r_{1}^{2}r_{2}^{2} \nonumber \\
& & + 10\,I_{1}R_{2}r_{2}^{4} - 10\,I_{3}R_{1}r_{1}^{4}r_{2}^{2} - 10\,I_{1}R_{3}r_{2}^{6} \nonumber \\
& & - 30\,I_{2}R_{2}r_{1}^{2}r_{2}^{4} + 5\,I_{4}R_{1}r_{1}^{6}r_{2}^{2} + 30\,I_{3}R_{2}r_{1}^{4}r_{2}^{4} \nonumber \\
& & + 30\,I_{2}R_{3}r_{1}^{2}r_{2}^{6} +  5\,I_{1}R_{4}r_{2}^{8} - I_{5}R_{1}r_{1}^{8}r_{2}^{2} \nonumber \\
& & - 10\,I_{4}R_{2}r_{1}^{6}r_{2}^{4} - 20\,I_{3}R_{3}r_{1}^{4}r_{2}^{6} - 10\,I_{2}R_{4}r_{1}^{2}r_{2}^{8} \nonumber \\
& & - I_{1}R_{5}r_{2}^{10} - \psi_{5}).
\end{eqnarray}
The coefficients $P$, $I_{m}$, $R_{m}$, and $\psi_{m}$ are defined (cf.\ Smith \& Theokas 1980) by the equations 
\begin{equation}
P(r_{1},r_{2},J,F) = \int_{0}^{min(r_{1},r_{2})} J(r) F(r) 2 \pi r dr,
\end{equation}
\begin{equation}
I_{m}(r_{1},J) = \int_{0}^{r_{1}} \frac{J(r)}{r_{1}^{2m-2}} \frac{\partial}{\partial r} (\pi
r^{2m}) dr,
\end{equation} 
\begin{equation}
R_{m}(r_{2},F) = \int_{0}^{r_{2}} \frac{F(s)}{r_{2}^{2m}} \frac{\partial}{\partial s} (s^{2m}) ds,
\end{equation}
and
\begin{eqnarray}
\psi_m & = & \psi_{m} (r_{1},r_{2},i,J,F) = \frac{L_1}{\pi\,r_1^2}\,\left(\frac{1 + 2\,m\,(r_1 + r_2)^2}{6\,(r_1 + r_2)^2}  \right. \nonumber \\
& & \times (\cos^2{i} - (r_2 - r_1)^2)^{3/2} - \sqrt{\cos^2{i} - (r_2 - r_1)^2} \nonumber \\
& & - \frac{m\,|r_2 - r_1|}{8\,(r_1 + r_2)^2}\,(\cos^4{i} - (r_2 - r_1)^4) \nonumber \\
& & + |r_2 - r_1|\,\left.\arctan{\left(\frac{\sqrt{\cos^2{i} - (r_2 - r_1)^2}}{|r_2 - r_1|}\right)} \right).
\end{eqnarray}

\subsection{The transparency and limb-darkening functions}
In our method, the transparency of the W-R wind is described by an analytical function that depends on a limited number of parameters. Since the functional form of the transparency is adopted {\it a priori}, our choice will obviously have a direct influence on the parameters derived for the system. Therefore, it is important to clearly specify the assumptions made in our approach. To avoid confusion with the standard symbols employed by Smith and Theokas (\cite{ST}), from now on, the various radii in our model will be denoted $\rho_{i}$ (i=1,2,3), where $\rho_1$ stands for the radius of the O-star that undergoes the eclipse. The use of the subscripts $e$ and $a$ will indicate whether the components are seen in emission or absorption.    

The first simplifying hypothesis is that all composite parts of the system are supposed to be spherically symmetrical. This means that the method is not applicable to close binary systems in which the components depart strongly from a spherical form as a result of tidal distortion and where {\it ellipticity} and {\it reflection} effects are both present.

\begin{figure}[h]
\begin{center}
\resizebox{8cm}{7.6cm}{\includegraphics{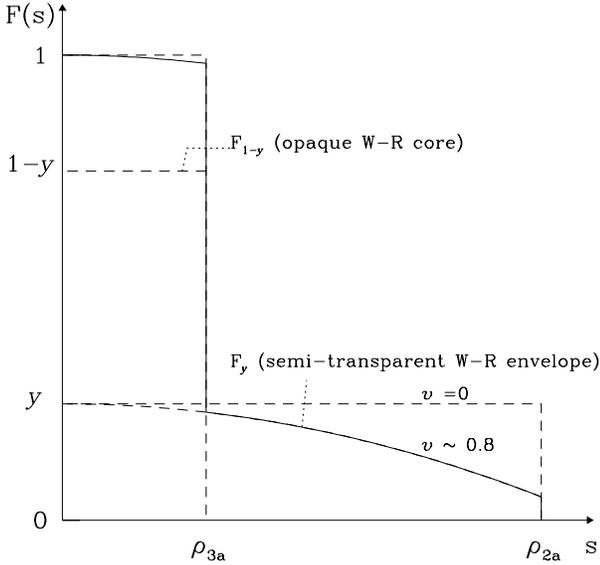}}
\end{center}
\caption{Schematical view of the transparency law across the disc of the W-R component as given by Eq.\,(\ref{eqn20}). The solid line shows the amount of light absorbed along the line of sight of impact parameter $s$. The individual contributions due to the opaque core and the semi-transparent extended atmosphere are illustrated.\label{fig-3}}
\end{figure}
Because a principle objective of the proposed technique of light curve analysis is the determination of the structure of a W-R envelope, a composite model consisting of an opaque core and a surrounding extended atmosphere was adopted for the W-R component. As a consequence, while for the transparency function $F(s)$ of the eclipsing W-R star, of radius $r_{0}$, Smith \& Theokas (\cite{ST}) simply adopted
\begin{equation}
F(s) = F_{y_a} (r_{0},\upsilon) = y_a \left[1 - \upsilon \label{eqn19} \left(\frac{s}{r_{0}}\right)^{2} \right] \hspace*{5mm} {\rm for} \hspace{2mm} s < r_{0},
\end{equation}
where $\upsilon$ is the coefficient of transparency, we chose as a first step a transparency law of the form
\begin{equation}
F(s) = F_{1-y_a} (\rho_{3a},0) + F_{y_a} (\rho_{2a},\upsilon) \label{eqn20},
\end{equation}
where the radius of the opaque core of the W-R star is $\rho_{3a}$ and that of the extended eclipsing envelope is $\rho_{2a}$ ($\geq \rho_{3a}$, see Fig.\ \ref{fig-3}). This transparency law was used in the preliminary analysis of the light curve of HD\,5980 by BP91 and is motivated by the fact that it corresponds to a physically more realistic model of the W-R star than that defined in Eq.\,(\ref{eqn19}), although the advantage of relative mathematical simplicity is still preserved.

For the brightness distribution $J(r)$ over the W-R disc (important for the eclipse of the W-R component by the O-star), a law very similar to that of the transparency function is adopted
\begin{equation}
J(r) = J(0) \left[ J_{1-y_e}(\rho_{3e},0) + J_{y_e}(\rho_{2e},u_2) \right] \label{eqn21},
\end{equation}
where $J(0)$ is the central surface brightness, $u_2$ is the coefficient of limb-darkening, and $J_{y_e}$ is defined as
\begin{equation}
J_{y_e}(r_{0},u) = y_e \left[ 1 - u \left(\frac{r}{r_{0}}\right)^2 \right] \label{eqn22}.
\end{equation}
When the W-R star is eclipsed, the radius of the core, assumed to be of uniform brightness, becomes $\rho_{3e}$ and that of the limb-darkened envelope $\rho_{2e}$. We note that the core and envelope radii of the W-R component seen in emission or absorption may differ.

We briefly address the physical meaning of this transparency law. First, in our model, the semi-transparent envelope of the W-R star has a finite extension given by the radii $\rho_{2a}$ and $\rho_{2e}$. However, the stellar winds of W-R stars do not stop abruptly so close to the star, but instead extend to large distances (much larger than the orbital separation in most binaries) to the shock with the interstellar (or circumstellar) medium. How can we then interpret $\rho_{2a}$ and $\rho_{2e}$? The radius $\rho_{2a}$ corresponds to the farthest position in the stellar wind where the residual optical depth along the line of sight produces a variation in the light curve that can be distinguished against the intrinsic photometric variability of the W-R star and the photometric errors. Similarly $\rho_{2e}$ is the outer radius of the W-R envelope that emits a measurable fraction of the light in the considered waveband. 

A clear difference between our approach and that of Antokhin \& Cherepashchuk (\cite{Igor}) concerns the functional behaviour of $F(s)$: whereas in our model, $F(s)$ is a convex function over the entire range $\rho \in [0, \rho_{3a}]$, Antokhin \& Cherepashchuk (\cite{Igor}) use a convexo-concave function, where the concave part corresponds to the stellar wind. As a consequence, $F(s)$ given by Eq.\,(\ref{eqn20}) decreases at a slower rate over the wind than the transparency law inferred by Antokhin \& Cherepashchuk. 

In contrast to Smith and Theokas (\cite{ST}) who neglect the effect, we take into account the limb-darkening of the OB-type component. Assuming that the formula employed by these authors to represent the brightness distribution across the W-R disc also applies to normal stars, we adopt the following limb-darkening law for the OB star
\begin{equation}
J(r) = \frac{L_{\rm O}}{\pi \rho_{1}^{2} (1 - u_1 + u_1^{2}/3)} \left[ 1 - u_1
\left(\frac{r}{\rho_{1}}\right)^{2} \right]^{2},
\end{equation}
where $L_{\rm O}$ is the luminosity of the OB star, of radius $\rho_{1}$, and $u_1$ is the coefficient of limb-darkening at the effective wavelength of the photometric filter considered.

Using these laws of transparency and limb-darkening, we then derived the expressions for $P$, $I_{m}$, $R_{m}$ (see Appendix\,\ref{app}), and $\psi_{m}$, and hence the final equations for the moments $B_{2m}$, corresponding to the primary and secondary minima.

\subsection{The orbital eccentricity}
The treatment of elliptical orbits in the frequency-domain was also addressed by Kopal (\cite{Kop2}). The problem still concerns the determination of the elements of the eclipse from the moments -- $B_{2m}$ in the present case -- derived from the light curve, but accounting for the eccentricity $e$ and the longitude of periastron $\omega$.

In the definition of the $B_{2m}$ moments, the phase-angle $\theta$ is no longer identical to the mean anomaly $M$ but has rather to be replaced by a linear function of the true anomaly $v$
\begin{equation}
\theta = v + \omega - \frac{\pi}{2}.
\end{equation}
The $d(\cos^{2m}{\theta})$ kernel in Eq.\,(\ref{eqn7}) thus becomes
\begin{equation}
d(\cos^{2m}{\theta}) = d \left[\sin^{2m}{(v+\omega)}\right].
\end{equation}
As a consequence, the empirical values of $B_{2m}$ cannot be derived from the observed data until a proper conversion of the phase angle into the true anomalies has been completed. This can be achieved either by a numerical inversion of Kepler's equation or the well-known asymptotic expansion of elliptical motion (e.g.\ Danjon \cite{danjon}, Kopal \& Al-Naimiy \cite{KAN})
\begin{eqnarray}
v & = & M + (2\,e - \frac{1}{4}\,e^{3})\sin{M} + (\frac{5}{4}\,e^{2} - \frac{11}{24}\,e^{4})\sin{2\,M} \nonumber \\
& & + \frac{13}{2}\,e^{3}\sin{3\,M} +
 \frac{103}{96}\,e^{4}\sin{4\,M} + ...
\end{eqnarray}

Regardless of the technique used to compute the true anomaly, this conversion evidently requires an {\it a priori} knowledge of $e$ as well as $\omega$. For a given value of the inclination $i$, these parameters can be derived by inversion of the equations (see e.g.\ Kopal \& Al-Naimiy \cite{KAN})
\begin{eqnarray}
\Delta\Phi & = & \frac{1}{2} + \frac{e\,\cos{\omega}}{\pi}\{1 + \csc^{2}{i} \nonumber \\
& & - \frac{e^{2}}{2}\,[\frac{8}{3}\,\cos^{2}{\omega} - 2 + O(\cot^2{i})]\}
\label{eqn-27},\end{eqnarray}
and
\begin{equation}
e\sin\omega = \frac{d_{2}-d_{1}}{4 \sin{\left( \frac{d_{1}+d_{2}}{4}\right)}}\,\left[1 - \frac{\cot^{2}{i}}{\sin^{2}{\left(\frac{d_{1}+d_{2}}{4}\right)}}\right]^{-1}
\label{eqn-28},
\end{equation}
where $\Delta\Phi$, $d_{1}$, and $d_{2}$ are, respectively, the phase displacement of the minima and the durations of the primary and secondary stellar core eclipses. These latter quantities are determined directly from the observed light curve. Since the orbital inclination of an eclipsing binary system is likely to be rather large, the $O(\cot^2{i})$ term in the coefficient of the $e^3$ term of Eq.\,(\ref{eqn-27}) can be neglected.

The empirical `elliptical' moments of the light curve then provide the elements of the binary exactly as in the `circular' case. One must ensure that the resulting values of the radii have been reduced to a constant unit of length. In our code, we therefore report all distances in relation to the semi-major axis $a$ of the relative orbit.

Our composite model adopted for the W-R star has different radii ($\rho_{2,3a}$ and $\rho_{2,3e}$) depending on whether the W-R component is seen as an eclipsing or an eclipsed disc. Because of this, although each individual half-eclipse provides an independent solution for the elements, the complete determination of the elements requires a combination of solutions obtained for both minima and because of the non-zero eccentricity, both the descending and ascending branches of each.

\section{Decoding of the light curve}
\subsection{Empirical determination of the moments}
We consider the light curve of a W-R binary system of eccentricity $e$ and period $P$, derived for a given photometric bandpass. The data are assumed to consist of a list of entries that provide for each observation the orbital phase $\Phi_{i}$ and the measured intensity $l_{i}$. To normalize the brightness scale, a mean intensity value is derived well outside the eclipses, during a phase-interval where the system is assumed to display (constant) maximum light. The $l_i$ value of each data point is then divided by this mean to normalize the light curve to unity. It has to be emphasized, however, that this does not necessarily imply that $L_{1} + L_{2} = 1$ for the W-R binary. The luminosity of a third photometrically unresolved component along the line of sight may indeed contribute to the observed brightness as well, thereby leading to a brightness distribution such as $L_{1} + L_{2} < 1$ in the final solution for the eclipse. In the case of an eccentric orbit, the eclipse-free mean $l_{i}$ value is preferably taken around apastron to avoid as much as possible any luminosity increase that could occur around periastron as a result of enhanced interaction effects between the components.\\

The determination of the moments $B_{2m}$ requires a smoothed light curve, which can be obtained, for instance, from a spline fit to the observed points with special attention to the minima. However, this task can become difficult if the descending or ascending branch of either minimum is ill-defined because of an uneven sampling of the observations or intrinsic photometric variability in the W-R star (see e.g.\ the case of WR\,22, Gosset et al.\ \cite{wr22}). A clustering of the points, in particular, is a serious handicap for the method. A preliminary processing performed, by filtering the observational data, to help reduce the scatter allows us then to obtain a smooth light curve $l(\Phi)$. 

The next step consists of determining the quantities $d_{1}$, $d_{2}$, and $\Delta\Phi$ (see above) from the smoothed light curve. For an assumed value of the orbital inclination $i$, the parameters $e$ and $\omega$ are obtained by means of an inversion of Eqs.\,(\ref{eqn-27}) and (\ref{eqn-28}). With these values of $e$ and $\omega$, the orbital phases $\Phi_{i}$ are converted into true anomalies.

The moments $B_{2m}$, which take into account the eccentricity effect, are obtained in practice by summation, using the following expression
\begin{eqnarray}
B_{2m} & = & \sum_{i=1}^{N-1} \left(\cos^{2m}{(\Theta_{i})} - \cos^{2m}{(\Theta_{i+1})}\right) \nonumber \\
& & \times \left(1 - \frac{l(\Theta_{i}) + l(\Theta_{i+1})}{2}\right),
\end{eqnarray}
where $\Theta_i$ are the predefined angles at which the smoothed light curve is sampled, $N$ is defined by the constant step $\Delta\Theta = \Theta_{i+1} - \Theta_{i}$ adopted, and the value of $\Theta_{1}$ corresponds to the first contact of the eclipse. The $l(\Theta_{i})$ values refer to the normalized smoothed light curve, and by definition $l = 1$ for $|\Theta| > |\Theta_{1}|$.

Since the individual data points are affected by observational errors, the integration of the empirical moments must itself be affected by errors. The uncertainty associated with the moments $B_{2m}$ can be evaluated using the following equation (see Al-Naimiy, \cite{AN1}, Smith \& Theokas \cite{ST})
\begin{eqnarray}
\Delta B_{2m} & = &\frac{1}{\sqrt{n}}\,\left\{\frac{1}{n}\,\sum_{j=1}^{n}
\left[l_{j} - l(\theta_{j})\right]^2\right\}^{1/2} \nonumber \\
& & \times \left(\cos^{2m}\theta_{1} - \cos^{2m}\theta_{n}\right),
\end{eqnarray}
where $n$ is the number of observed points over the considered eclipse, and $l_{j} - l(\theta_{j})$ is the difference between the observed point of index $j$ and the smoothed light curve at $\theta_j$. The angles $\theta_{1}$ and $\theta_{n}$ refer, respectively, to the first and last observed data point over the relevant part of the light curve.

\subsection{Solution for the elements}
For each half-eclipse, there are five non-linear algebraic equations to be solved simultaneously for the elements $\rho_{1}$, $\rho_{2a}$ or $\rho_{2e}$, $\rho_{3a}$ or $\rho_{3e}$, $i$, $u_{1}$ or $u_{2}$, $y_{a}$ or $y_{e}$, $\upsilon$, $L_{1}$ or $L_{2}$ the meanings of which are summarized below for convenience:

$L_{1}$ = luminosity of the OB-type star,

$L_{2}$ = luminosity of the W-R star,

$i$ = inclination angle of the orbit,

$\rho_{1}$ = radius of the OB-type star,

$\rho_{2a,e}$ = radius of the W-R envelope seen in absorption or emission,

$\rho_{3a,e}$ = radius of the W-R opaque core seen in absorption or emission,

$u_{1}$ = limb-darkening coefficient of the OB-type star,

$u_{2}$ = limb-darkening coefficient of the W-R envelope,

$\upsilon$ = transparency coefficient of the W-R envelope (cf. Fig.\ \ref{fig-3}), and

$y_{a,e}$ = contribution of the W-R envelope in absorption or emission (cf. Fig. \ref{fig-3})
 
This large number of variables can fortunately always be reduced to a smaller number for both the primary and secondary minima, as explained hereafter. The preliminary determination of the orbit inclination $i$ already eliminates for instance one variable.

At the primary minimum, when the OB star is in front, according to our composite model $\upsilon \equiv 0$ by definition. Since we must also have that $0 \leq u_{2} \leq 1$, solutions can be searched for a set of discrete values of the parameter $u_{2}$ in this interval, so that the remaining variables are $\rho_{1}$, $\rho_{2e}$, $\rho_{3e}$, $y_{e}$, and $L_{2}$. 

At the secondary minimum, when the W-R star eclipses the OB component, $u_{1}$ is the limb-darkening coefficient of the OB star. The value of $u_1$ can be adopted following e.g.\ the tabulated values supplied by Klinglesmith \& Sobieski (\cite{KS}). According to our model, we now have $0 \leq \upsilon \leq 1$, so that again, after selection of a sample of $\upsilon$ values, we can proceed in solving the equations for the remaining parameters $\rho_{1}$, $\rho_{2a}$, $\rho_{3a}$, $y_{a}$, and $L_{1}$ only.

The solution of the system of as many as five non-linear equations $$B_{2m}(\rho_{1}, \rho_{2a}, \rho_{3a}, i, u_1, y_a, \upsilon, L_1, L_2) = B_{2m}({\rm observed})$$ or $$B_{2m}(\rho_{1}, \rho_{2e}, \rho_{3e}, i, u_2, y_e, \upsilon, L_1, L_2) = B_{2m}({\rm observed})$$
is obtained by minimizing the $\chi^2$ compiled from the residuals of these equations
\begin{equation}
\chi^2 = \sum_{m=1}^N \frac{|B_{2m}({\rm computed}) - B_{2m}({\rm observed})|^2}{\Delta\,B_{2m}^2}.
\end{equation} 
This minimization is achieved by means of Powell's technique (e.g., Press et al.\ \cite{NumRec} and references therein).

\section{Application to HD\,5980}
\subsection{HD\,5980: a peculiar system}
HD\,5980 $\equiv$ AB\,5 (Azzopardi \& Breysacher \cite{Azzo79}) is associated with  NGC\,346, the largest H\,{\sc ii} region + OB star cluster in the Small Magellanic Cloud. This remarkable W-R binary, which underwent an LBV-type event in August 1994, is presently recognized as a key-object for improving our understanding of massive star evolution. HD\,5980 is a rather complex system because it consists of at least three stars: two stars form the eclipsing binary with the 19.266\,day period, whilst the third component, an O-star, which is detected by means of a set of absorption lines and by means of its third light (see also below), could be a member of a highly eccentric 96.5\,day period binary (Schweickhardt \cite{Schweick}, Foellmi et al.\ \cite{Foellmi}). Whether or not the third star is physically bound to the eclipsing binary remains currently unclear. Before the LBV eruption, both components of the eclipsing binary already showed emission lines in their spectra and were thus classified as Wolf-Rayet stars (Niemela \cite{Niemela}). However, as shown by the analysis of the spectra taken during and after the LBV event, at least the star that underwent the eruption (hereafter called star A) was not a classical, helium-burning, Wolf-Rayet object, but rather a WNha star, i.e., a rather massive star with substantial amounts of hydrogen present in its outer layers (Foellmi et al.\ \cite{Foellmi}). These WNha stars have wind properties that are intermediate between those of extreme Of stars and classical WN stars.

A summary of the light changes exhibited by HD\,5980 was presented by Breysacher (\cite{Brey97}). The technique of light curve analysis described above is applied to HD\,5980 prior to the outburst. Given that star A, the component in front of its companion (hereafter called star B) during primary eclipse, was a WNha star and since there are no indications of wind effects in the primary eclipse, we assume that this component behaves as an OB-star in the light curve. 705 measurements with the Stroemgren $v$ filter, described in BP91, are taken into consideration. The shape of the resulting light curve does not allow us to use the ill-defined ascending branch of the primary eclipse (star A in front) for the analysis. Therefore, only three half-minima will be considered. Compared to the preliminary analysis carried out by BP91, the software tool presently used has been upgraded, allowing us for instance to assess the quality of the solution by means of a synthetic light curve.

\subsection{Solutions of the light curve}
From the smoothed light curve, the durations of the primary and secondary stellar core eclipses as well as the separation between the core eclipses are measured first to be $d_1 = 0.062 \pm 0.005$,  $d_2 = 0.095 \pm 0.005$, and $\Delta\,\Phi = 0.362$. All durations are expressed as phase intervals (i.e.\ fractions of the orbital cycle). The corresponding values of the eccentricity $e$ and the longitude of periastron $\omega$ are: $e = 0.314 \pm 0.007$ and $\omega = 132.5^{\circ} \pm 1.5^{\circ}$. These values are in fairly good agreement with those derived for these parameters by Breysacher \& Fran\c{c}ois (\cite{BF}) ($e = 0.30 \pm 0.02$, $\omega = 135^{\circ} \pm 10^{\circ}$) from a completely different approach based on the analysis of the width variation of the He {\sc ii} $\lambda$\,4686 line using the analytical colliding-wind model devised by L\"{u}hrs (\cite{LU}). From radial-velocity studies, Kaufer et al. (\cite{Kau02}) and Niemela et al.\ (\cite{Niemela2}) also found that $e = 0.297 \pm 0.036$ and $e = 0.28$, respectively. The values of the moments of the light curve of HD\,5980 are listed in Table\,\ref{moments}. 

The inclination angle of the orbit can easily be derived with reasonable accuracy. For each of the three half-minima, a quick analysis is carried out for a number of plausible discrete values of $i$ ($i = 82^{\circ}, 83^{\circ}, ... 89^{\circ}$), and the value of $i$ finally adopted is the one for which closest agreement is obtained between the three solutions provided for the radius of star A and the radius of the opaque core of star B. These quantities are fundamental elements of the system and the combination of the solutions of the descending and ascending branches of both minima must indeed provide, at the end of the detailed analysis, a unique value for $\rho_{1}$ and very similar - if not identical - values for $\rho_{3a}$ and $\rho_{3e}$. The above conditions are fullfilled for $85^{\circ} \leq i \leq 87^{\circ}$, therefore we adopt $i = 86^{\circ} \pm 1^{\circ}$.

\begin{table}
\caption{Values of the moments of the eclipses of HD\,5980 used throughout this paper. These values are derived from the light curve prior to the 1994 LBV event. \label{moments}}
\begin{tabular}{l c c c}
\hline
& Primary eclipse & \multicolumn{2}{c}{Secondary eclipse} \\
\cline{3-4}
& Ingress & Ingress & Egress \\ 
\hline\hline
B$_2$  & $.00935 \pm .00026$ & $.00832 \pm .00041$ & $.00481 \pm .00024$ \\
B$_4$  & $.01801 \pm .00048$ & $.01542 \pm .00065$ & $.00922 \pm .00043$ \\
B$_6$  & $.02604 \pm .00067$ & $.02164 \pm .00080$ & $.01326 \pm .00059$ \\
B$_8$  & $.03351 \pm .00083$ & $.02721 \pm .00088$ & $.01699 \pm .00071$ \\
B$_{10}$ & $.04046 \pm .00097$ & $.03226 \pm .00094$ & $.02044 \pm .00082$ \\ 
\hline
\end{tabular}
\end{table}

The second step in the procedure consists of solving the equations for $i = 86^{\circ}$, each half-eclipse being treated in a completely independent manner. We recall that all radii are reduced to the semi-major axis of the relative orbit. 

We first consider the descending branch of the primary eclipse. Solutions for $\rho_{1}, \rho_{2e}, \rho_{3e}, L_{2}$, and $y_{e}$ are searched for discrete values of the parameter $u_{2}$ (0.1, 0.2, 0.3, ... 1) with rather broad variation ranges allowed to the variable parameters. Each search sequence is based on a series of 1000 trials. A first set of results is obtained to provide convergence rates (i.e., an estimate of the likelihood of the derived solutions) and a $\sigma$ value for each parameter. A second iteration with reduced variation ranges (the original one $\pm \sigma$) for the parameters leads to a second set of solutions with improved convergence rates and lower $\sigma$ values. The process is repeated five times, until a stabilization of the parameter values becomes noticeable. For the last iteration, the grid of $u_{2}$ values is enlarged significantly and the respective convergence rates of the corresponding solutions are used to determine the best fit $u_{2}$ value and to estimate the error on this parameter. The obtained results are $\rho_{1} = 0.150 \pm 0.004$, $\rho_{2e} = 0.257 \pm 0.018$, $\rho_{3e} = 0.110 \pm 0.005$, $L_{2} = 0.300 \pm 0.016$, $u_{2} = 0.58 \pm 0.07$, and $y_{e} = 0.19 \pm 0.03$.  
            
For the secondary eclipse, solutions are searched for $\rho_1$, $\rho_{2a}$, $\rho_{3a}$, $L_1$, and $y_a$. The choice of the parameter $u_1$ for the limb-drakening law (Klinglesmith \& Sobieski \cite{KS}) of star A has little impact on the other parameters. We repeated the fitting procedure for different values of $u_1$ (0.1, 0.3, 0.5, 0.7) and recovered the same solutions within the errorbars. The largest sensitivity was found for $\rho_{2a}$ when $u_1 = 0.7$. In this rather unlikely case, $\rho_{2a}$ exceeds its usual value by $1.5\,\sigma$. In the following, we thus focus on the results obtained with $u_1 = 0.3$, which seems a reasonable value for star A (Klinglesmith \& Sobieski \cite{KS}). The descending and ascending branches are analysed separately and solutions are searched for discrete values of the parameter $\upsilon$ (0.1, 0.2, 0.3, ... 1). The same iterative procedure as described above for the primary eclipse is applied. The resulting solutions for the secondary descending branch are $\rho_{1} = 0.160 \pm 0.006$, $\rho_{2a} = 0.259 \pm 0.018$, $\rho_{3a} = 0.110 \pm 0.005$, $L_{1} = 0.385 \pm 0.028$, $y_{a} = 0.19 \pm 0.03$, and $\upsilon = 0.60 \pm 0.20$; and for the secondary ascending branch  $\rho_{1} = 0.163 \pm 0.008$, $\rho_{2a} = 0.290 \pm 0.019$, $\rho_{3a} = 0.105 \pm 0.003$, $L_{1} = 0.412 \pm 0.026$, $y_{a} = 0.20 \pm 0.03$, and $\upsilon = 0.45 \pm 0.15$        
       
The differences between the solutions provided by the three half-minima are relatively small compared to the errors and average values are derived for the parameters. A distinction between {\it absorption} and {\it emission} values does not appear to be necessary any longer. The error in the mean for each parameter $p_{i}$ is computed in a conservative approach using the expression 
\begin{equation}     
\sigma = \left( \frac{\sum_{i=1}^n(p_i - <p>)^2 + \sum_{i=1}^n \sigma_{i}^2}{nN} \right)^{1/2}\label{eqn-33},
\end{equation} 
where n is the number of values used in the mean and N the number of independent sets (half-eclipses) of values. For an analysis completed in this way, the values adopted for the parameters of the stellar components in the HD\,5980 binary system are given in Table\,\ref{solutions}.

\begin{figure*}[htb]
\begin{center}
\resizebox{16cm}{!}{\includegraphics{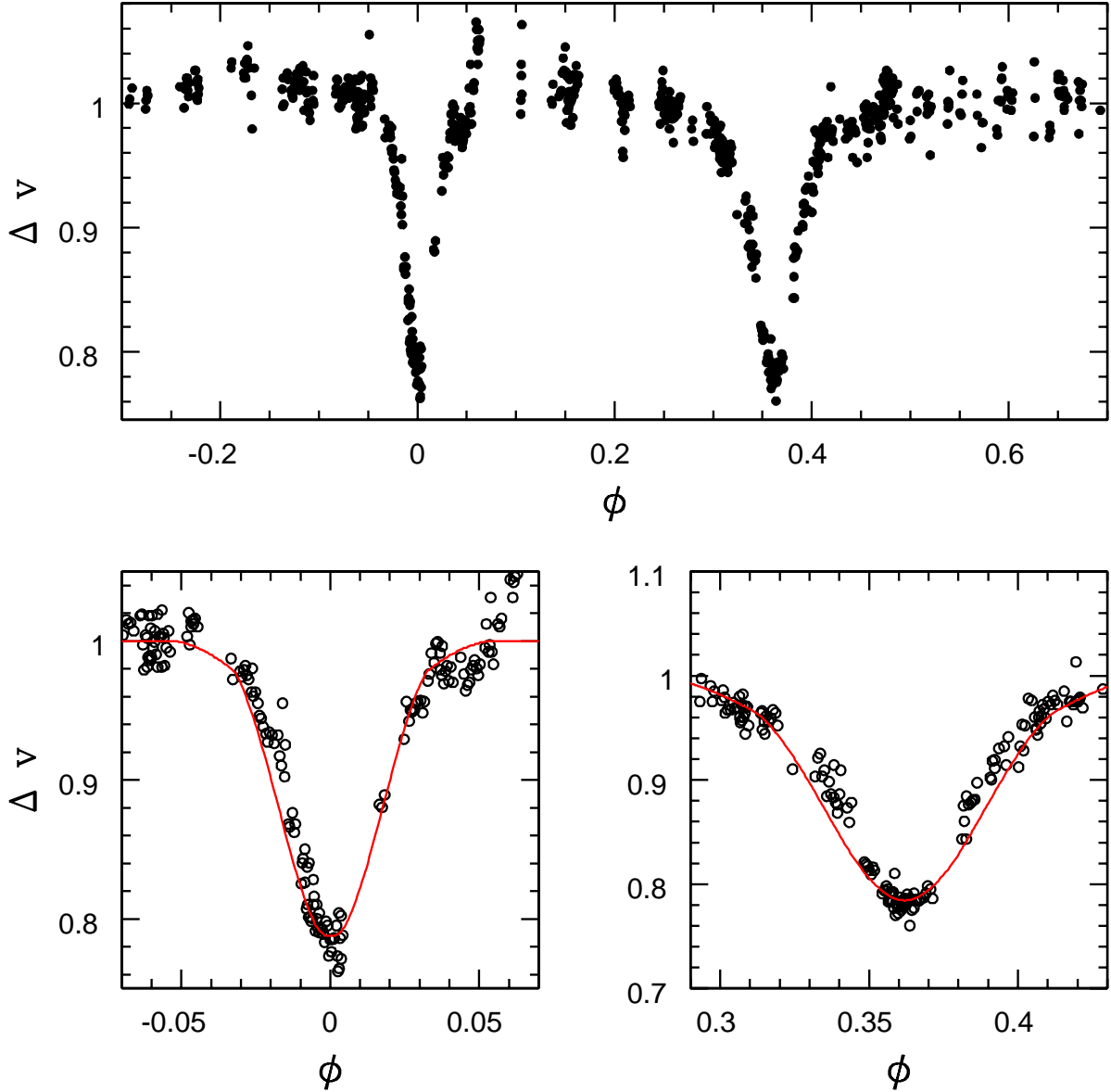}}
\end{center}
\caption{Top panel: the observed light curve of HD\,5980 in the Stroemgren $v$ filter as a function of orbital phase. The lower left and lower right panels show the synthetic light curves for the primary and secondary eclipses respectively compared to the actual data. The synthetic light curves were computed using the mean parameters of the solutions found by our programme.
 \label{fig-4}}
\end{figure*}

\begin{table}
\caption{Final `best-fit' values of the model parameters of the stars in the HD\,5980 binary system for $i = 86^{\circ}$ and prior to the 1994 LBV outburst.\label{solutions}}
\begin{tabular}{l c c c}
\hline
& Primary eclipse & \multicolumn{2}{c}{Secondary eclipse} \\
\cline{3-4}
& Ingress & \multicolumn{2}{c}{Ingress \& Egress} \\ 
\hline\hline
star A  &  \multicolumn{3}{c}{$\rho_1 = 0.158 \pm 0.005$} \\
        &  & \multicolumn{2}{c}{$L_1 = 0.398 \pm 0.021$} \\
\hline
star B  & \multicolumn{3}{c}{$\rho_3 = 0.108 \pm 0.003$}\\
        & \multicolumn{3}{c}{$\rho_2 = 0.269 \pm 0.014$}\\
        & $L_2 = 0.300 \pm 0.016$ & & \\
        & \multicolumn{3}{c}{$y = 0.19 \pm 0.02$} \\
        & $u_2 = 0.58 \pm 0.07$ & \multicolumn{2}{c}{$v = 0.52 \pm 0.14$} \\
\hline
\end{tabular}
\end{table}

\subsection{Discussion}
Figure \ref{fig-4} shows how the synthetic light curve derived from the above elements fits the observational data for both the primary and secondary eclipses. While the bottom and the wings are fairly well fitted, the transparency law that we adopted for the extended envelope of star B is probably still too crude to allow a perfect match to the observations of the descending and ascending sides of both minima. As a consequence, the size of this envelope is probably slightly overestimated by our model. An asymmetry in this envelope, inferred by BP91, is difficult to ascertain from the present study. The difference between the values of $\rho_{2a}$ provided by the two branches of the secondary eclipse is indeed marginally significant only given the errors. $L_1 + L_2 \neq 1$ is independent confirmation of an unresolved source of third light along the line of sight that accounts for an additional relative luminosity of $L_{3} = 0.302$. In principle, one would expect a superior control of the uniqueness of the solution if the brightness ratios $L_2/L_1$ and $L_3/L_1$ in the optical could be fixed independently of our light curve analysis. However, in the specific case of HD\,5980, it is impossible to infer these brightness ratios from spectroscopy. In this system, there have been changes in both the spectroscopic and in photometric properties of the binary components, and the brightness ratios are thus epoch dependent. Spectrophotometric brightness ratios would have to be estimated based on the dilution of the emission lines. However, as shown by recent spectroscopic studies (e.g.\ Foellmi et al.\ \cite{Foellmi}), these emission lines contain epoch-dependent contributions from components A and B, as well as from the wind-wind interaction region. In our analysis of the light curve, we thus preferred not to constrain the brightness ratios {\it a priori}. Our results can however be checked {\it a posteriori} against the results obtained in the UV domain. Assuming that an O7 supergiant, which does not partake in the orbit, was responsible for this third light, Koenigsberger et al.\ (\cite{Ko94}) estimated a value of 2.8 for the ratio $(L_{1}+L_{2})/L_{3}$, which was considered to be consistent with the 2.0 obtained from the BP91 elements. It is noticeable that the value of 2.3 derived from the present analysis is in even closer agreement with the estimate of Koenigsberger et al.\ (\cite{Ko94}).

\begin{figure*}[t!hb]
\begin{minipage}{6cm}
\begin{center}
\resizebox{6cm}{!}{\includegraphics{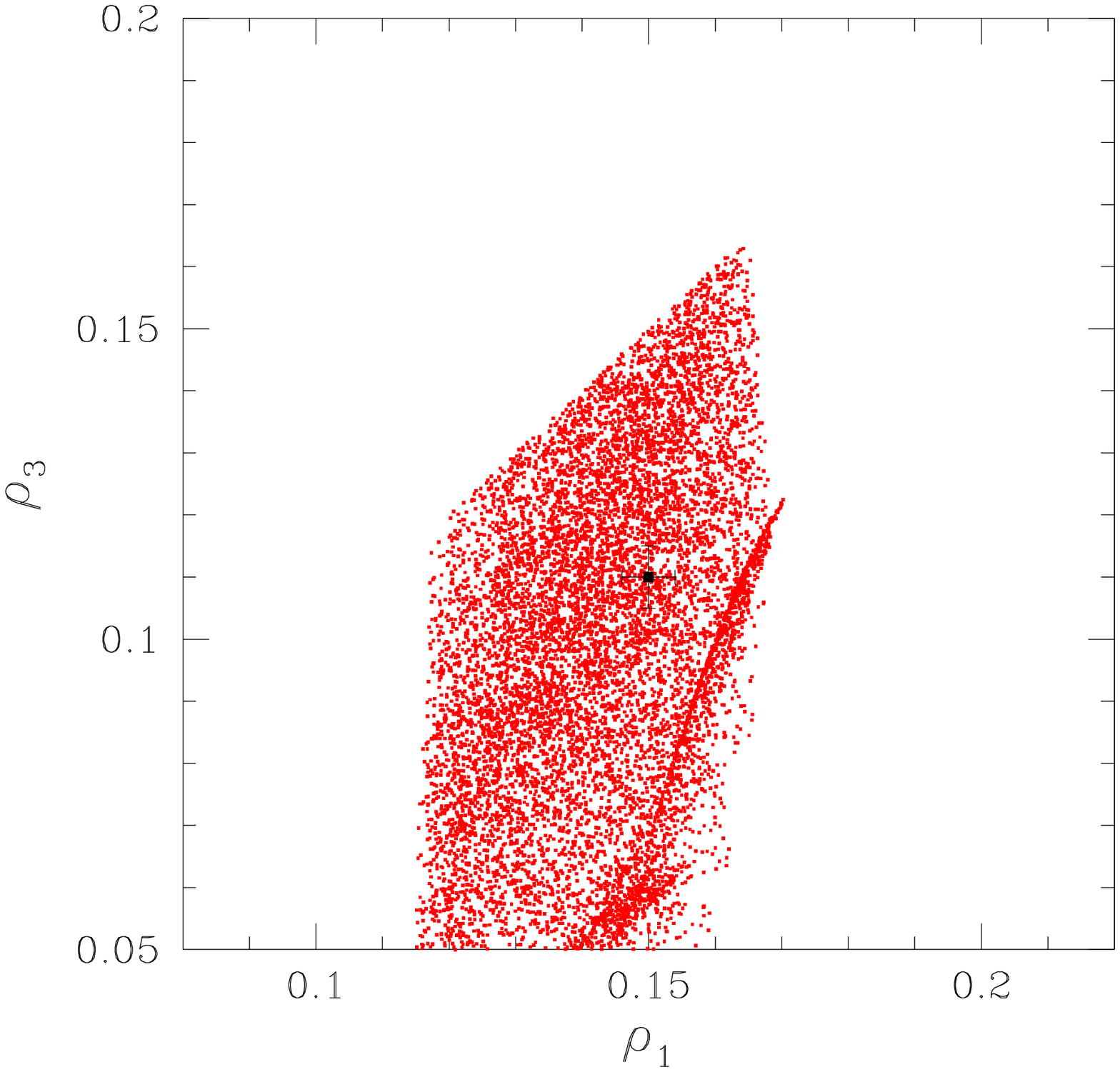}}
\end{center}
\end{minipage}
\begin{minipage}{6cm}
\begin{center}
\resizebox{6cm}{!}{\includegraphics{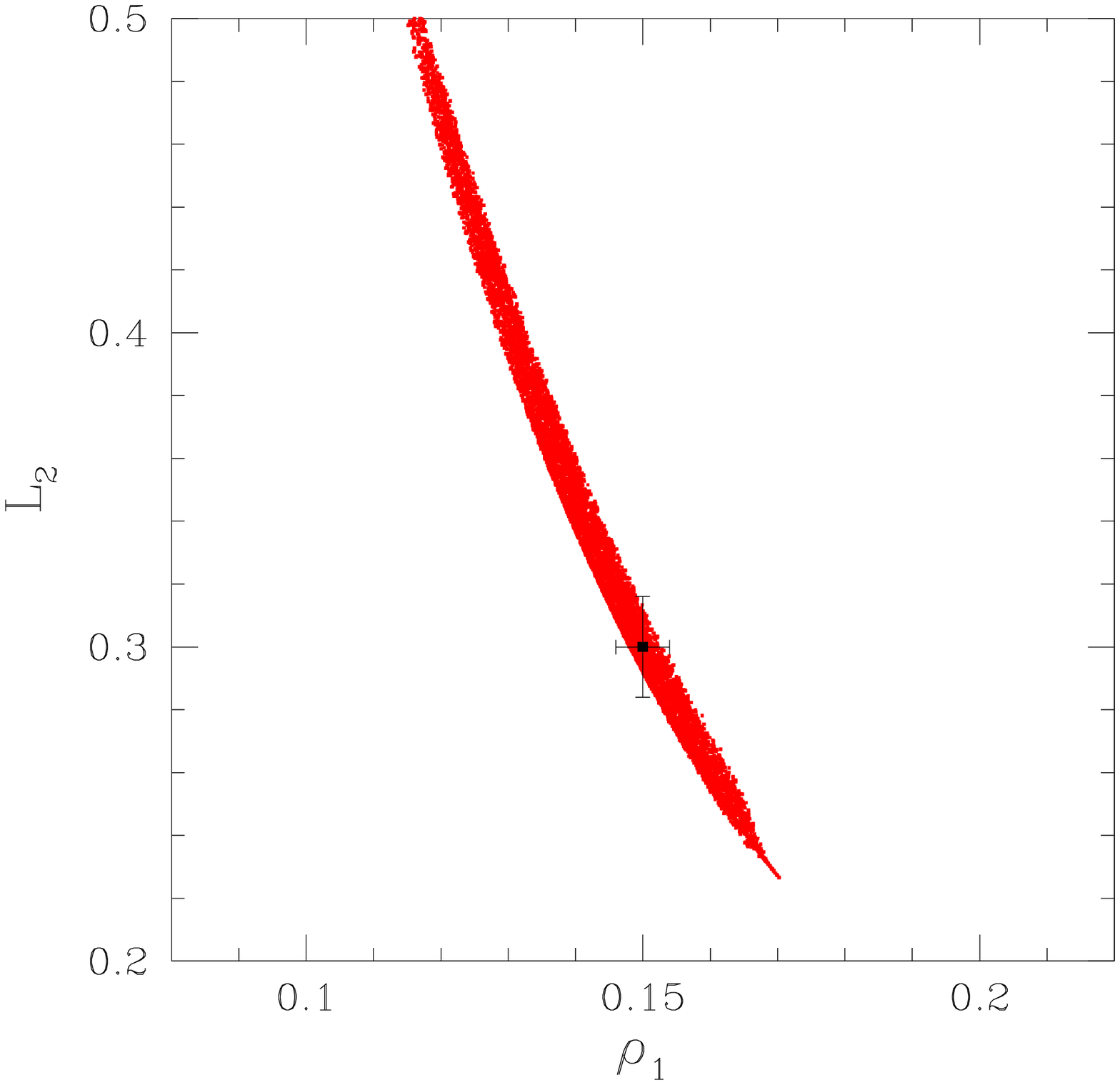}}
\end{center}
\end{minipage}
\begin{minipage}{6cm}
\begin{center}
\resizebox{6cm}{!}{\includegraphics{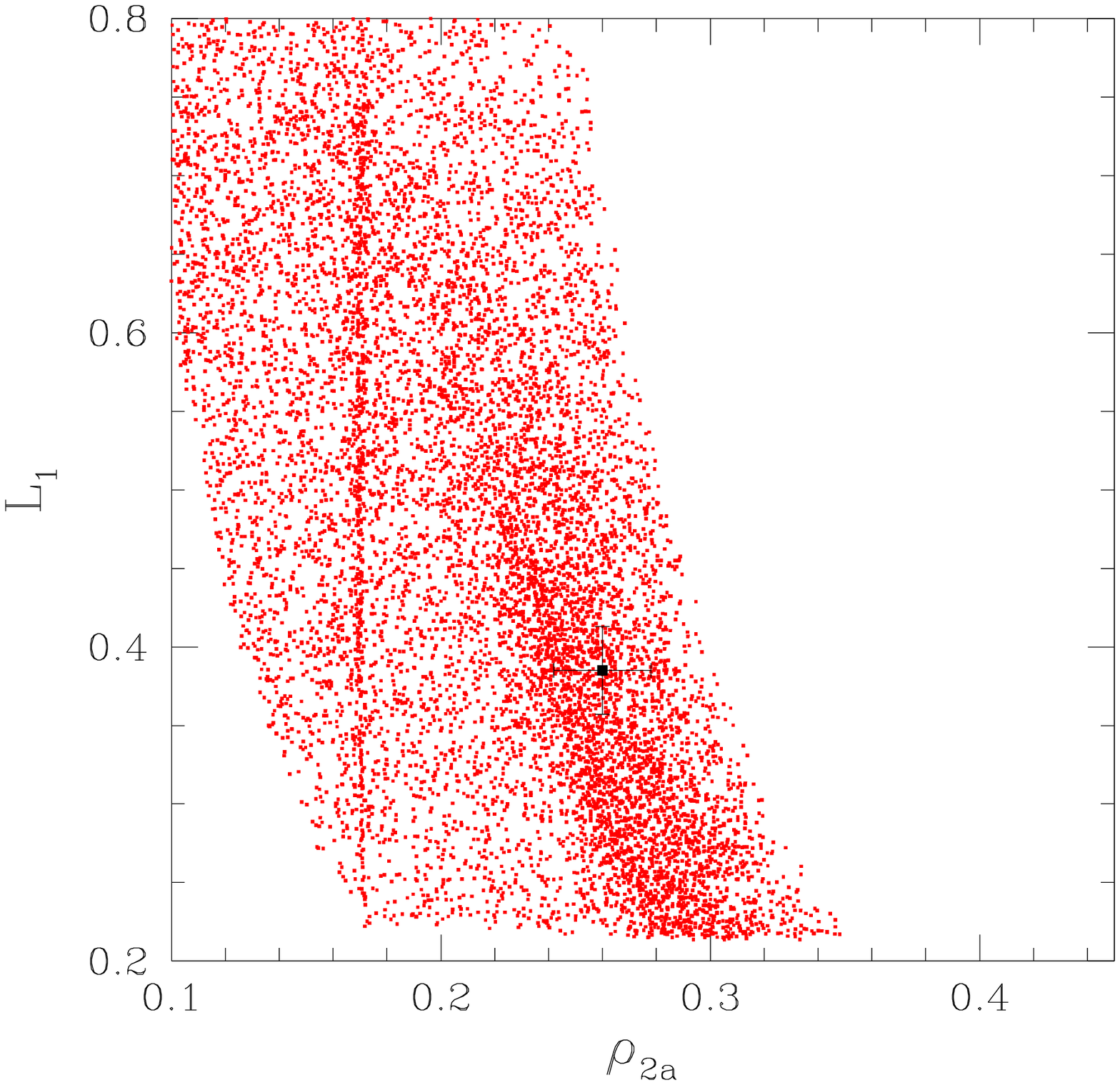}}
\end{center}
\end{minipage}
\caption{Distribution of the parameters derived from the inversion of the moments of the light curve in various parameter planes. The left and middle panels are derived from the primary eclipse, whilst the rightmost panel presents the solutions of the secondary eclipse. The final solutions derived from each eclipse as well as their $1-\sigma$ error bars are overplotted in each panel. \label{fig-locus}}
\end{figure*}
Adopting $M_{v} = -7.5$ for the global absolute visual magnitude of HD\,5980 (cf. Koenigsberger et al.\ \cite{Ko94}), the magnitudes of the binary components are $M_{v}(star A) = -6.50$ and $M_{v}(star B) = -6.19$. Because of the difficulties in assigning the various lines of the spectrum of HD\,5980 to a specific component, there have been few attempts to establish a full SB2 orbital solution (Niemela et al.\ \cite{Niemela2}, Foellmi et al.\ \cite{Foellmi}). Therefore, our knowledge of the component masses of the eclipsing binary remains uncomfortably poor, even though we are dealing with an eclipsing system. Foellmi et al.\ (\cite{Foellmi}) used a multi-component fit of the N\,{\sc iv} $\lambda$\,4058 and N\,{\sc v} $\lambda$\,4603 lines to derive absolute masses of 58 -- 79\,M$_{\odot}$ for star A and 51 -- 67\,M$_{\odot}$ for star B. These values differ significantly from the estimates of Niemela et al.\ (\cite{Niemela2}), who attributed the entire N\,{\sc iv} $\lambda$\,4058 line to star A and the entire N\,{\sc v} $\lambda$\,4603 emission to star B when inferring minimum masses of $m\,\sin^3{i} = 28$ and 50\,M$_{\odot}$, respectively. Adopting the solution of Foellmi et al.\ (\cite{Foellmi}), the sum of the masses of the components of the eclipsing binary equals 109 --  146\,M$_{\odot}$. This corresponds to a semi-major axis of 0.67 -- 0.74\,AU (144 -- 159\,R$_{\odot}$) for the binary orbit, and to stellar radii of 22.7 -- 25.1\,R$_{\odot}$ for star A, 15.6 -- 17.2\,R$_{\odot}$ for the core of star B, and 38.7 -- 42.8\, R$_{\odot}$ for its envelope.    

Alternatively, one could estimate absolute masses from the visual magnitudes evaluated above by means of mass-luminosity relations derived e.g., from massive star evolutionary models. However, this approach is hampered by a lack of precise knowledge of the spectral types of the stars (and hence their bolometric corrections) and by the fact that the components of HD\,5980 can probably not be considered as `normal' early-type stars. 
   
We considered the correlations between the various free parameters by plotting them in pairs (see Fig.\,\ref{fig-locus}). For this purpose, we used the results of 10\,000 trials for each eclipse. In most cases, we do not observe obvious correlations; the solutions are scattered over a limited part of the parameter plane (see e.g., the radius of the core of star B, $\rho_3$, versus that of star A, $\rho_1$, derived from the primary eclipse). However, the solutions show a clear trend, if we plot the luminosity of star A ($L_1$) as a function of the radius of the envelope of star B ($\rho_{2a}$, evaluated from the secondary eclipse), and fall even along a clearly defined locus if we plot the luminosity of star B ($L_2$) as a function of the radius of star A ($\rho_1$, derived from the primary eclipse). These trends can be understood at least qualitatively. In fact, for the primary eclipse (the opaque core of star A occulting star B), an increase in the radius of star A implies that a larger fraction of the light of star B is removed. To account for the observed depth of the light curve, the total luminosity of star B must decrease. On the other hand, during secondary eclipse (star B in front of star A), an increase in the radius of the semi-transparent envelope of star B will produce deeper and broader wings of the secondary eclipse. To account for the observed eclipse shape, the model must react in terms of an increase in the surface brightness (and hence the luminosity) of star A. 

In principle, the photometric variability of HD\,5980 as detected through medium-band filter observations could be affected by the Doppler shift of strong emission lines that fall in the wavelength range covered by the photometric filters. To quantify this effect, we simulated an observation of the star through the ESO Stroemgren filters. For this purpose, we used the spectrum of HD\,5980 taken from the spectrophotometric catalogue of Morris et al.\ (\cite{Morris}) kindly provided to us by Dr.\ J.-M.\ Vreux. Our calculations indicate that less than 1\% of the flux in the $v$ band comes from emission lines. Therefore, we do not expect the Doppler shift to have any significant effect on the photometric variability discussed here. The situation would be quite different, if we were dealing with data taken in the $b$ band. There, about 30\% of the flux is produced by emission lines (especially the strong He\,{\sc ii} $\lambda$\,4686 line). In this case, the Doppler motion as well as variations in the line intensity caused by the wind-wind interaction (e.g., Breysacher et al.\ \cite{BMN}, Breysacher \& Fran\c cois \cite{BF}) could lead to significant variations in the observed $b$ magnitude.
\begin{figure}[htb]
\begin{center}
\resizebox{8cm}{!}{\includegraphics{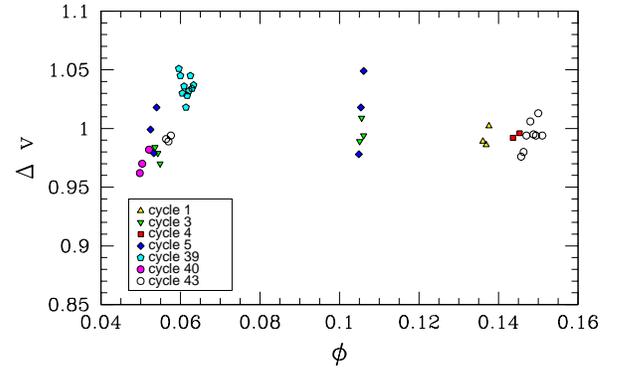}}
\end{center}
\caption{Observed $v$ magnitudes of HD\,5980 in the phase interval 0.04 -- 0.16. The various symbols indicate data from different orbital cycles. 
Periastron passage occurs at phase 0.061.\label{fig-peri}}
\end{figure}

Finally, we note that the data points of the light curve suggest an increase in the brightness of the HD\,5980 system at orbital phases after $\phi \sim 0.04$ (see Fig.\,\ref{fig-4}). Whilst this increase could be caused by light reflections from the wind interaction region (which should have its concavity roughly turned towards the observer) near periastron, we caution that this apparent trend is actually inferred from observations from a single campaign (cycle 39 in Fig.\ \ref{fig-peri}) and could therefore be related to intrinsic variability of the WR star rather than represent a genuine phase-locked effect. 

\section{Conclusions}
We have presented a method that allows us to treat the problem of atmospheric eclipses in the light curves of moderately wide, eccentric Wolf-Rayet + O binary systems in a semi-analytical way. We have then applied this method to the light curve of the peculiar system HD\,5980 prior to its 1994 outburst. Despite the non-uniform sampling of the light curve, the fact that we are analysing data from different orbital cycles (hence affected differently by the intrinsic variability of the WR star) and the limitations of our assumptions on the properties of the WR envelope, our method yields consistent results when applied to the primary or secondary eclipse. We have been able to constrain some of the physical parameters of this system, although the lack of a consistent SB2 spectroscopic orbital solution prevents us from obtaining fully model independent parameters. As a next step, we will try to generalize our method to the analysis of the eclipses of a system harbouring two stars, both with extended atmospheres. This should allow us to analyse the light curve of HD\,5980 observed after the LBV eruption.

\appendix
\section{Analytical expressions for the moments of the eclipses \label{app}}
\subsection{Primary eclipse (OB star in front)}
Here the relevant radii are: $\rho_1$, the radius of the OB-star; $\rho_{2e}$ and $\rho_{3e}$, the radii of the W-R core and envelope in emission. We note that we have assumed that $\rho_{2e} \leq \rho_1$. Adopting a surface brightness law for the W-R star given by Eqs.\,(\ref{eqn21}) and (\ref{eqn22}) and $F(s) = 1$ (for $s \leq \rho_1$), we find that its total luminosity is given by
\begin{equation}
L_{\rm W-R} = J(0)\,\pi\,\rho_{2e}^2\,\left[(1 - y_e)\,\left(\frac{\rho_{3e}}{\rho_{2e}}\right)^2 + y_e\,\left(1 - \frac{u_2}{2}\right)\right],
\end{equation}
\begin{equation}
I_m = L_{\rm W-R}\,\frac{(1 - y_e)\,\left(\frac{\rho_{3e}}{\rho_{2e}}\right)^{2m} + y_e\,\left(1 - \frac{m\,u_2}{m + 1}\right)}{(1 - y_e)\,\left(\frac{\rho_{3e}}{\rho_{2e}}\right)^{2} + y_e\,\left(1 - \frac{u_2}{2}\right)},
\end{equation}
and
\begin{equation}
P = L_{\rm W-R}\,\frac{(1 - y_e)\,\left(\frac{\rho_{3e}}{\rho_{2e}}\right)^2 + y_e\,g\,[1 - \frac{u_2}{2}\,g]}{(1 - y_e)\,\left(\frac{\rho_{3e}}{\rho_{2e}}\right)^2 + y_e\,\left(1 - \frac{u_2}{2}\right)},
\end{equation}
where $g(\rho_1, \rho_{2e}) = \min{\left(1, \left(\frac{\rho_1}{\rho_{2e}}\right)^2\right)}$. The expression of $P$ reduces to 
\begin{equation}
P = L_{\rm W-R} \hspace*{3mm} {\rm if} \hspace*{3mm} \rho_{2e} < \rho_1,
\end{equation}
and finally,
\begin{equation}
R_m = 1.
\end{equation}
\subsection{Secondary eclipse (W-R star in front)}
Here $\rho_{2a}$ and $\rho_{3a}$ are the radii of the W-R core and envelope in absorption, respectively. Again, we have assumed that $\rho_{2a} \leq \rho_1$. Adopting a transparency law for the W-R star given by Eqs.\,(\ref{eqn19}) and (\ref{eqn20}), we find that
\begin{equation}
L_{\rm O} = J(0)\,\pi\,\rho_{1}^2\,\left(1 - u_1 + \frac{u_1^2}{3}\right), \end{equation}
\begin{equation}
I_m = \frac{L_{\rm O}}{1 - u_1 - \frac{u_1^2}{3}}\,\left(1 - \frac{2\,u_1\,m}{m + 1} + \frac{u_1^2\,m}{m + 2}\right),
\end{equation}
and
\begin{eqnarray}
P & = &\frac{L_{\rm O}}{1 - u_1 + \frac{u_1^2}{3}}\,\left\{(1 - y_a)\,\left(\frac{\rho^2_{3a}}{\rho^2_{1}} - u_1\,\frac{\rho^4_{3a}}{\rho^4_1} + \frac{u_1^2}{3}\,\frac{\rho^6_{3a}}{\rho^6_1}\right)\right. \nonumber \\ 
& + & y_a\,h\,\left[\left(1 - u_1\,h + \frac{u_1^2\,h^2}{3}\right)\right. - \upsilon\,g\,\left(\frac{1}{2} - \frac{2\,u_1\,h}{3} \right. \nonumber \\
& + & \left.\left.\left. \frac{u_1^2\,h^2}{4}\right)\right]\right\},
\end{eqnarray}
where $g(\rho_1, \rho_{2a}) = \min{\left(1, \left(\frac{\rho_1}{\rho_{2a}}\right)^2\right)}$ and $h(\rho_1, \rho_{2a}) = 
\min{\left(1, \left(\frac{\rho_{2a}}{\rho_{1}}\right)^2\right)}$.\\
Finally, we obtain 
\begin{equation}
R_m = (1 - y_a)\,\left(\frac{\rho_{3a}}{\rho_{2a}}\right)^{2m} + y_a\,\left(1 - \frac{m\,\upsilon}{m + 1}\right)
\end{equation}

\acknowledgement{We are greatly indebted to Drs.\ S.A.H.\ Smith and A.C.\ Theokas for clarification about their method, and Profs.\ G.\ Koenigsberger and A.\ Maeder for fruitful discussions. GR is supported by the FRS/FNRS (Belgium) and through the XMM/INTEGRAL PRODEX contract (Belgian Federal Science Policy Office) as well as by the Communaut\'e Fran\c caise de Belgique - Action de recherche concert\'ee - Acad\'emie Wallonie - Europe.}

\end{document}